\documentclass[fleqn,usenatbib]{mnras}

\usepackage[T1]{fontenc}

\DeclareRobustCommand{\VAN}[3]{#2}
\let\VANthebibliography\thebibliography
\def\thebibliography{\DeclareRobustCommand{\VAN}[3]{##3}\VANthebibliography}

\usepackage{graphicx}	
\usepackage{amsmath}	
\usepackage{amssymb}	

\usepackage[normalem]{ulem}
\usepackage{xcolor}

\usepackage{savesym}
\savesymbol{tablenum}

\usepackage[binary-units]{siunitx}
\restoresymbol{SIX}{tablenum}

\renewcommand\arcsec{\mbox{$^{\prime\prime}$}}%

\defcitealias{smercina22}{S22}

\title[ALMA CO\,(2--1) Observation of PSBs]{{Do Post-Starburst Galaxies Host Compact Molecular Gas {Reservoirs}?}}
\author[F.\ Sun \& E.\ Egami]{
Fengwu Sun$^{1}$\thanks{E-mail: fengwusun@email.arizona.edu}
and Eiichi Egami$^{1}$ 
\\
$^{1}$Steward Observatory, University of Arizona, 933 N. Cherry Avenue, Tucson, 85721, USA 
}

\date{Accepted 2022 October 11. Received 2022 October 4; in original form 2022 June 17}

\pubyear{2022}

\usepackage{newtxtext,newtxmath}

\begin{document}
\label{firstpage}
\pagerange{\pageref{firstpage}--\pageref{lastpage}}
\maketitle

\begin{abstract}
We analysed the high-resolution (up to $\sim$\,0\farcs2) ALMA CO\,(2--1) and 1.3\,mm dust continuum data of eight gas-rich post-starburst galaxies (PSBs) in the local Universe, six of which had been studied by Smercina et al.\ (2022).  
In contrast to this study reporting the detections of extraordinarily compact {(i.e., unresolved)} reservoirs of molecular gas in the six PSBs, 
our visibility-plane analysis resolves the CO\,(2--1) emission in all eight PSBs with effective radii ($R_\mathrm{e,CO}$) of $0.8_{-0.4}^{+0.9}$\,kpc, typically consisting of gaseous components at both circumnuclear and extended disc scales. 
With this new analysis, we find that the CO sizes of gas-rich PSBs are {compact with respect to their stellar sizes (median ratio $=0.43_{-0.21}^{+0.27}$), {but comparable to the sizes of the gas discs}} seen in local luminous infrared galaxies (LIRGs) and early-type galaxies.
We also find that the CO-to-stellar size ratio of gas-rich PSBs is {potentially} correlated with the gas depletion time scale, placing them as transitional objects between LIRGs and early-type galaxies from an evolutionary perspective.  
Finally, the star formation efficiency of the observed PSBs appear consistent with those of star-forming galaxies on the Kennicutt-Schmidt relation, showing no sign of suppressed star formation from turbulent heating.
\end{abstract}

\begin{keywords}
galaxies: evolution -- galaxies: ISM -- galaxies: star formation
\vspace{-8pt}
\end{keywords}


\vspace{-18pt}

\section{Introduction}
\label{sec:01_intro}

Post-starburst galaxies (PSBs), or "E+A" galaxies, are recognised as the galaxy class undergoing a rapid transition from vigorously star-forming galaxies (e.g., luminous infrared galaxies, LIRGs) to quiescent early-type galaxies (see a recent review by \citealt{french21}).
The optical spectra of these galaxies are dominated by A-type stars formed $10^{2}-10^{3}$\,Myr ago but without strong nebular emission lines, indicating a low level of on-going star-formation rate (SFR).
Given the rapid decline of SFR over a time scale of $\sim10^2$\,Myr, one would expect a quick depletion of molecular gas reservoirs through star formation, stellar and active galactic nucleus (AGN) feedback \citep[e.g.,][]{hopkins06,sell14}.
However, significant amounts of molecular gas reservoirs have been found in many PSBs \citep[e.g.,][]{french15,rowlands15,baron22}, placing them {below} the global Kennicutt-Schmidt relation \citep{kennicutt98} between the surface densities of gas mass and SFR in certain studies \citep[e.g.,][]{li19}. 
These discoveries raise a critical question on how the star-forming activities are suppressed in the PSB environments.

\citet[][hereafter \citetalias{smercina22}]{smercina22} reported very compact CO\,(2--1) emission (median $R_\mathrm{e,CO} < 0.2$\,kpc) in six local gas-rich PSBs, $\sim$10\% of the stellar continuum sizes and  consistent with those observed for CO\,(3--2) lines in \citet{french22}.
If true, this would further suggest a high turbulent heating in the interstellar medium (ISM) of PSBs, which could suppress the star-forming efficiency by up to an order of magnitude.
In this work, we perform a re-analysis of the archival ALMA CO\,(2--1) and dust continuum observations for eight gas-rich PSBs, six of which were initially studied by \citetalias{smercina22}. 
The observations and corresponding data reduction techniques are described in Section~\ref{sec:02_obs}.
The measurements of CO\,(2--1) and dust continuum properties are presented in Section~\ref{sec:03_res}.
In Section~\ref{sec:04_dis}, we compare our results with \citetalias{smercina22}, and discuss the implication of gas size measurements and star-forming activities for PSBs in our sample. 
The conclusions are summarised in Section~\ref{sec:05_sum}.
Throughout this paper, we assume a flat $\Lambda$CDM cosmology with $h=0.7$ and $\Omega_m = 0.3$.

\vspace{-18pt}

\section{Observations and Data Reduction}
\label{sec:02_obs}

The ALMA CO\,(2--1) and 1.3\,mm dust continuum observations of eight gas-rich PSBs were obtained in Cycle 3 and 4 through Programme 2015.1.00665.S and 2016.1.00980.S (PI: Smith), using both the 12\,m array and compact 7\,m array.
Among them, six sources have been studied by \citetalias{smercina22}, and the observation setups are detailed there.
Two sources (0413 and 2276) observed with these ALMA programmes were not presented in \citetalias{smercina22}.
0413 was only observed with the 7\,m array with a 1.3\,hr on-source integration;
2276 was observed with the 12\,m array for 14\,min and 7\,m array for 28\,min on-source.
The minimum, median and maximum of the baseline lengths of the ALMA observations are listed in Table~\ref{tab:01_all}.

We processed the ALMA data with \textsc{casa} \citep{casa} using the versions with the Cycle\,3/4 pipelines (v4.3.1--v4.7.2).
The 12\,m and 7\,m data were concatenated for imaging and \textit{uv}-plane analysis.
We performed continuum imaging with the line-free channels and line-imaging with the \textit{uv}-continuum-subtracted calibrated measurement sets.
Both the continuum and CO\,(2--1) line were imaged at the native resolution using the Briggs weighting (\textsc{robust}$=$0.5) {without further \textit{uv}-tapering}, resulting in a range of angular resolutions from 0\farcs18 (0480) to 5\arcsec\ (0413).
In order to inspect the possible existence of extended emission and produce images with uniform angular resolutions, we also employed \textit{uv}-tapering techniques with kernel sizes of 20, 40, 80 and 250\,\si{\kilo\lambda} (\textsc{robust}$=$2.0), corresponding to synthesised beam sizes from 5\arcsec\ (i.e., 7\,m-array-like beam sizes) to 0\farcs8.
For the line-imaging, we adopted a velocity bin size of 10\,\si{km.s^{-1}}.

\begin{figure*}
\centering
\includegraphics[width=0.49\linewidth]{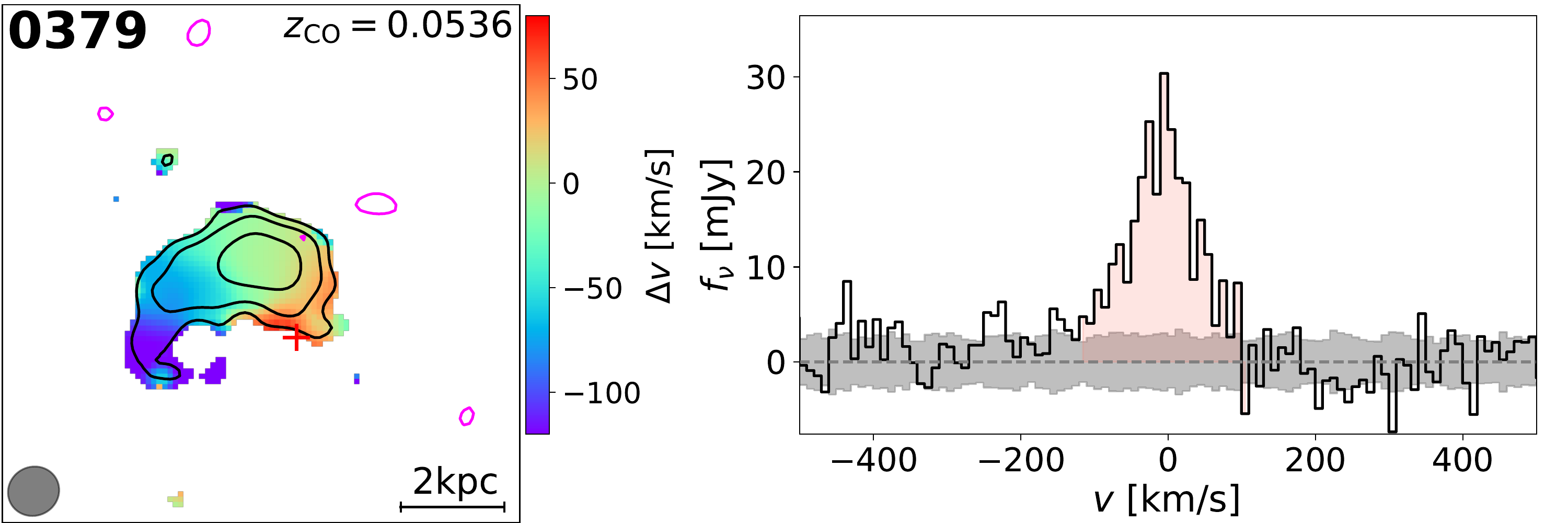}
\includegraphics[width=0.49\linewidth]{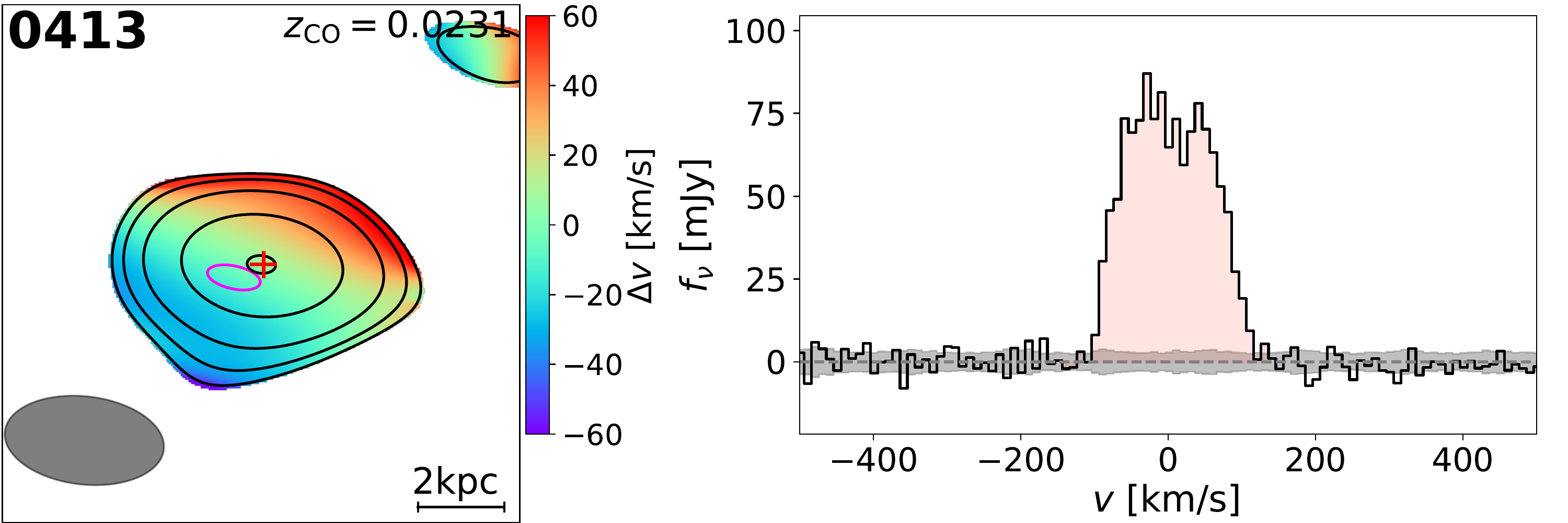}
\includegraphics[width=0.49\linewidth]{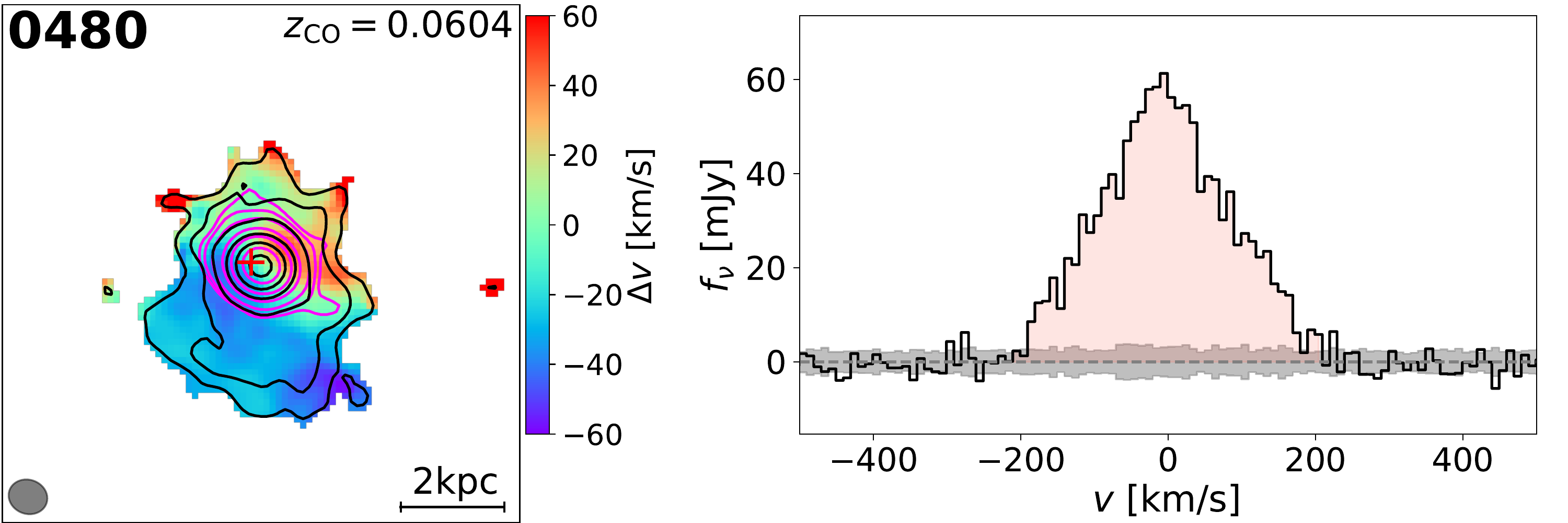}
\includegraphics[width=0.49\linewidth]{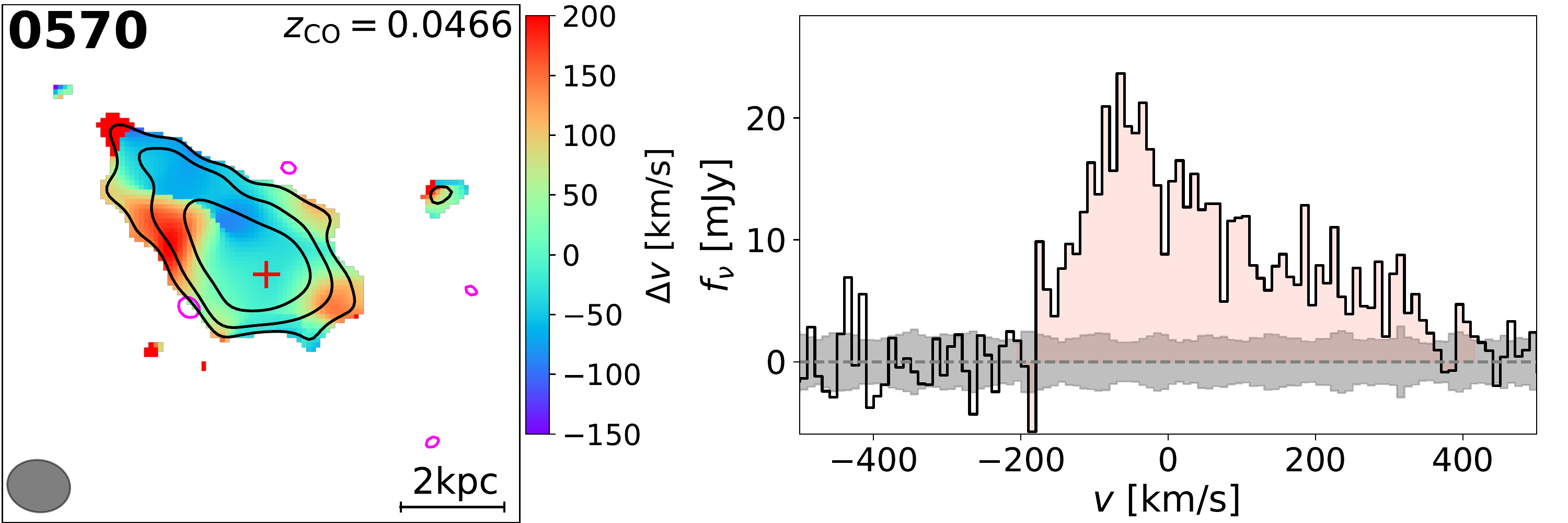}
\includegraphics[width=0.49\linewidth]{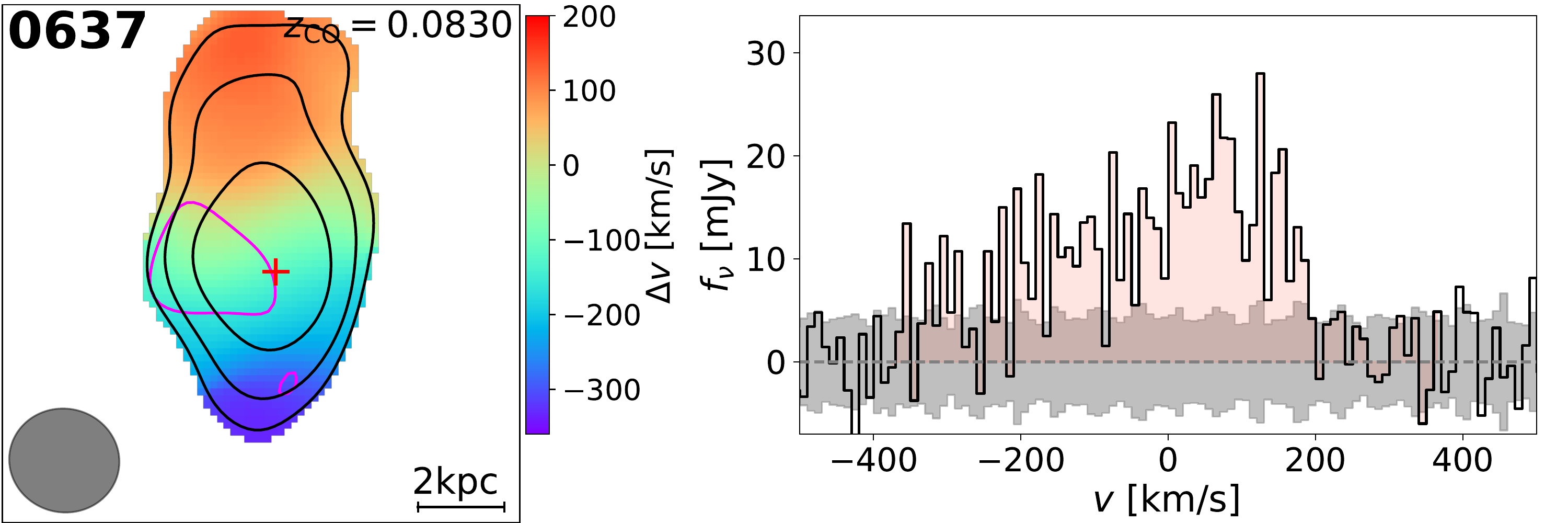}
\includegraphics[width=0.49\linewidth]{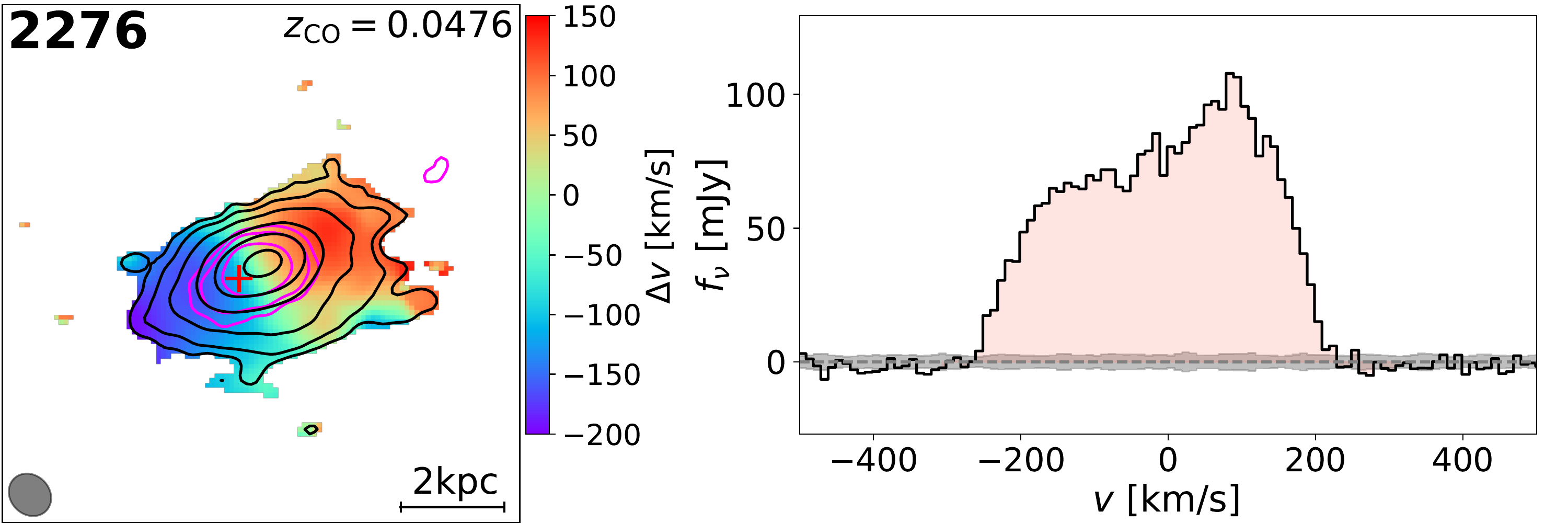}
\includegraphics[width=0.49\linewidth]{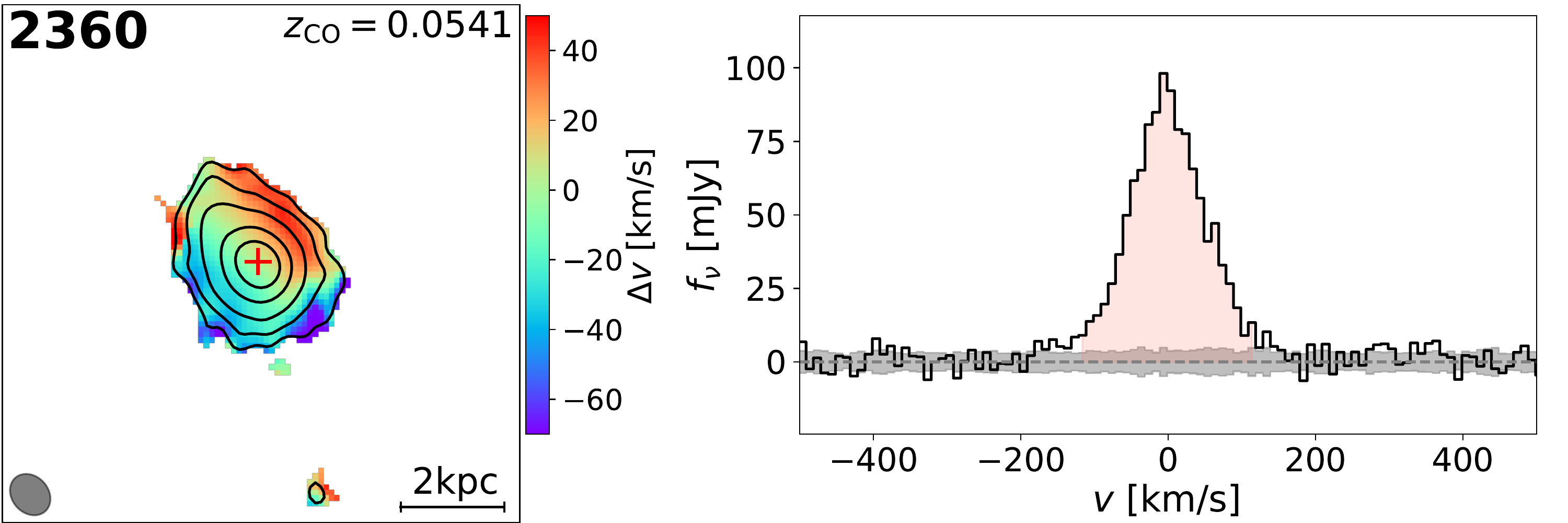}
\includegraphics[width=0.49\linewidth]{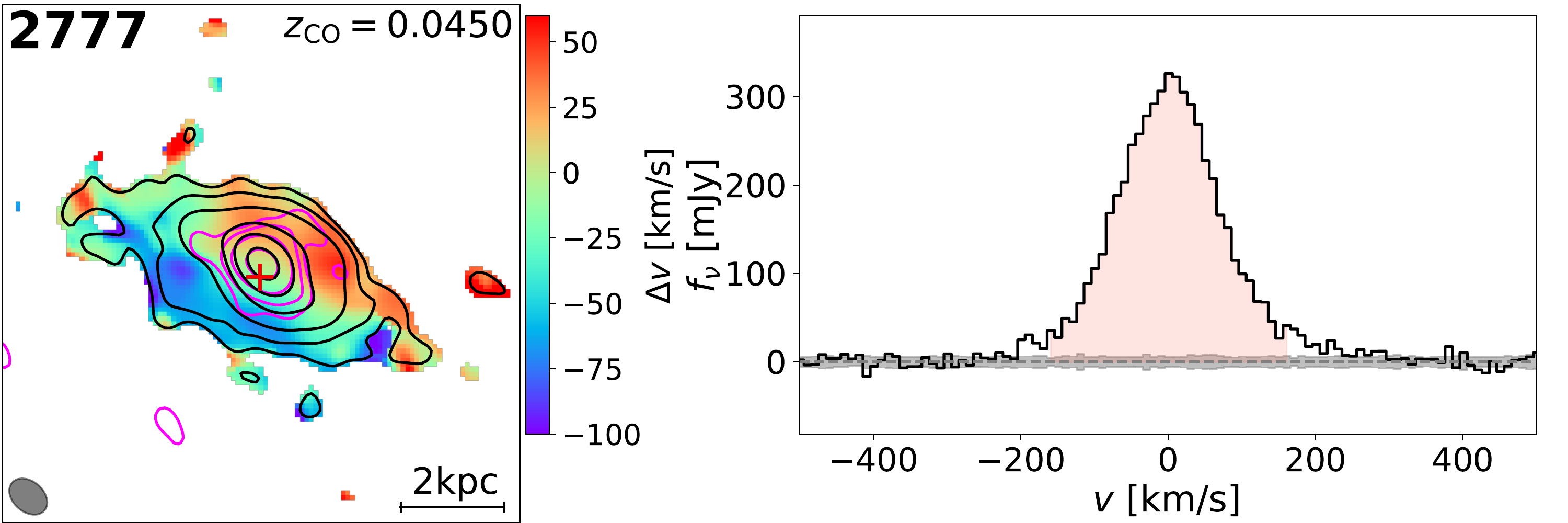}
\caption{Morphology and kinematics of CO\,(2--1) emission in eight gas-rich PSBs.
For each source we show their CO\,(2--1) moment\,1 (intensity-weighted velocity field) map on the left, overlaid with the contours of moment\,0 (integrated intensity) map at levels of 3, 5, 10, 25, 50, 100$\sigma$ from the outside in (solid black lines). 
Magenta contours denote the 1.3\,mm continuum maps at the same significance levels (detected for five sources), and red crosses denote the optical centres measured in the SDSS $r$-band.
Synthesised beam sizes are plotted as the grey ellipses in the bottom-left corners.
The CO\,(2-1) spectra extracted at the peak pixel of the lowest-resolution ($20k\lambda$-tapered; {in which the sources are unresolved}) line cubes are shown on the right.
The RMS noise is shown with the filled grey area, and the velocity ranges used for the moment\,0 map integration are shown in shallow red.
}
\vspace{-8pt}
\label{fig:01_spec}
\end{figure*}

\vspace{-12pt}

\section{Results}
\label{sec:03_res} 

\begin{figure*}
\centering
\includegraphics[width=\linewidth]{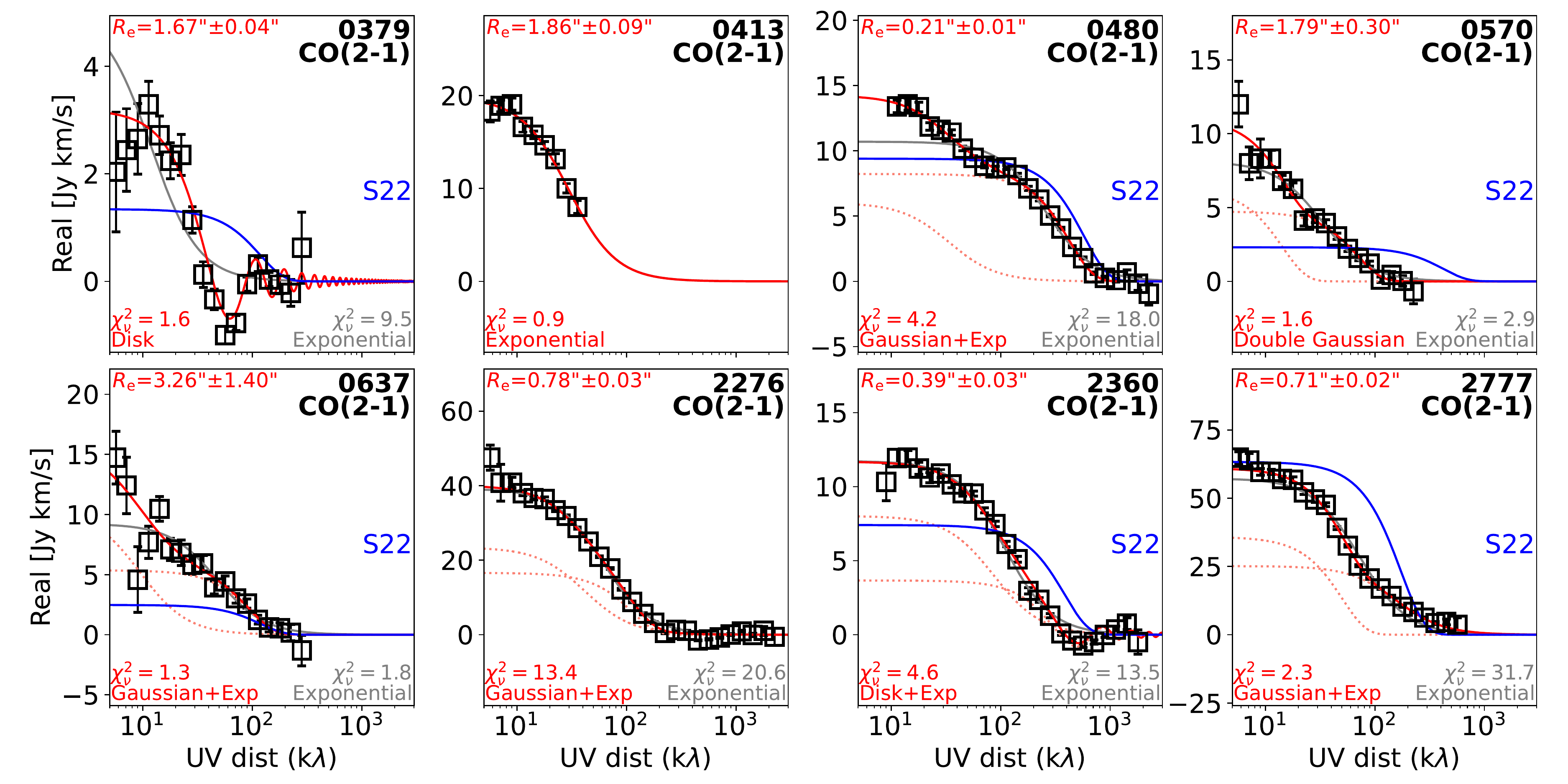}
\vspace{-10pt}
\caption{Visibility profiles of CO\,(2--1) emission detected with eight gas-rich PSBs (black squares).
In each plot, the best-fit structural profile model is shown as the solid red line, with the model name and reduced $\chi^2$ noted in the lower-left corner.
If multiple components are involved, the profile of each component is shown as dotted red line.
Circularised effective radius of CO\,(2--1) emission is noted in the top-left corner of each panel.
Best-fit single exponential disc profiles are shown as the solid grey curves, and single Gaussian profiles based on \citetalias{smercina22} measurements are shown in blue curves for comparison.
}
\vspace{-8pt}
\label{fig:02_vis}
\end{figure*}

\begin{table*}
\centering
\caption{Properties of PSBs in our sample.}
\label{tab:01_all}
\begin{tabular}{cccccccccc} 
\hline
ID & $z_\mathrm{CO}$ & R.A. & Decl. & Baselines & $S_\mathrm{1.3mm}$ & $S\Delta v_\mathrm{CO(2-1)}$ & $L^\prime_\mathrm{CO(2-1)}$ & $R_\mathrm{e,CO}$ & $R_\mathrm{e,SDSS}$ \\
& & & & (\si{k\lambda}) & (\si{\milli Jy}) & (\si{Jy.km.s^{-1}}) & (\si{10^8.K.km.s^{-1}.pc^{2}}) & (\si{\kilo pc}) &  (\si{\kilo pc}) \\
(1) & (2) & (3) & (4) & (5) & (6) & (7) & (8) & (9) & (10) \\
\hline
0379 & 0.0536 & 22:55:06.88 & $+$00:58:39.4 &    6--93--289 &        $<$0.19 &  3.2$\pm$0.2 &  1.06$\pm$0.06 & 1.72$\pm$0.04 & 1.87$\pm$0.03 \\
0413 & 0.0231 & 03:16:54.98 & $-$00:02:31.6 &     6--17--36 &  0.23$\pm$0.10 & 19.8$\pm$0.4 &  1.20$\pm$0.03 & 0.86$\pm$0.04 & 2.03$\pm$0.04 \\
0480 & 0.0604 & 09:48:18.68 & $+$02:30:04.1 & 10--321--2279 &  4.11$\pm$0.05 & 14.2$\pm$0.4 &  6.00$\pm$0.16 & 0.25$\pm$0.01 & 1.89$\pm$0.03 \\
0570 & 0.0466 & 09:44:26.99 & $+$04:29:57.3 &    6--67--243 &        $<$0.19 & 11.0$\pm$1.0 &  2.75$\pm$0.26 & 1.65$\pm$0.25 & 2.58$\pm$0.05 \\
0637 & 0.0830 & 21:05:08.67 & $-$05:23:59.7 &    6--88--321 &  0.86$\pm$0.25 & 15.7$\pm$4.2 & 12.68$\pm$3.39 & 5.08$\pm$2.41 & 7.18$\pm$0.31 \\
2276 & 0.0476 & 08:34:33.63 & $+$17:20:43.5 &  6--276--2307 &  1.02$\pm$0.08 & 40.0$\pm$1.6 & 10.44$\pm$0.42 & 0.74$\pm$0.03 & 1.66$\pm$0.05 \\
2360 & 0.0541 & 09:26:19.30 & $+$18:40:40.9 & 11--341--2293 &        $<$0.11 & 11.7$\pm$0.4 &  3.94$\pm$0.13 & 0.41$\pm$0.03 & 1.89$\pm$0.13 \\
2777 & 0.0450 & 14:48:16.03 & $+$17:33:05.9 &   6--169--638 &  4.72$\pm$0.24 & 61.0$\pm$1.3 & 14.16$\pm$0.31 & 0.62$\pm$0.02 & 3.00$\pm$0.05 \\
\hline
\end{tabular}
\vskip1mm\hskip0mm\begin{minipage}{\linewidth} 
{{\bf Notes:} 
Columns: (1) Source ID as the SDSS plane number (same as that in \citealt{smercina18}). Except for 0413 and 2276, the other six sources have been analysed in \citetalias{smercina22}. 
(2) Redshift of CO\,(2-1) peak (Figure~\ref{fig:01_spec}) measured by single Gaussian fitting; 
(3)--(4) Right Ascension and Declination of CO\,(2--1) peak; 
(5) Minimum, median and maximum of ALMA baseline lengths at the redshifted CO wavelength (unit: \si{\kilo\lambda}). Baseline lengths provided by 7\,m array are at $6\sim36$\,\si{\kilo\lambda}, and 12\,m array provides baselines with lengths of $11\sim2000$\,\si{\kilo\lambda};
(6) 1.3\,mm continuum flux density. If undetected,  $3\sigma$ upper limit is measured from the continuum map whose beam size has the smallest difference from the CO size;
(7) CO\,(2--1) line flux modelled on the \textit{uv}-plane;
(8) CO\,(2--1) line luminosity computed from Column (2) and (7);
(9) Circularised effective radius of CO\,(2--1) emission measured on the \textit{uv}-plane;
(10) Circularised effective radius of SDSS $r$-band image measured with \textsc{galfit}.
}\end{minipage}
\vspace{-8pt}
\end{table*}

\begin{figure*}
\centering
\includegraphics[width=0.49\linewidth]{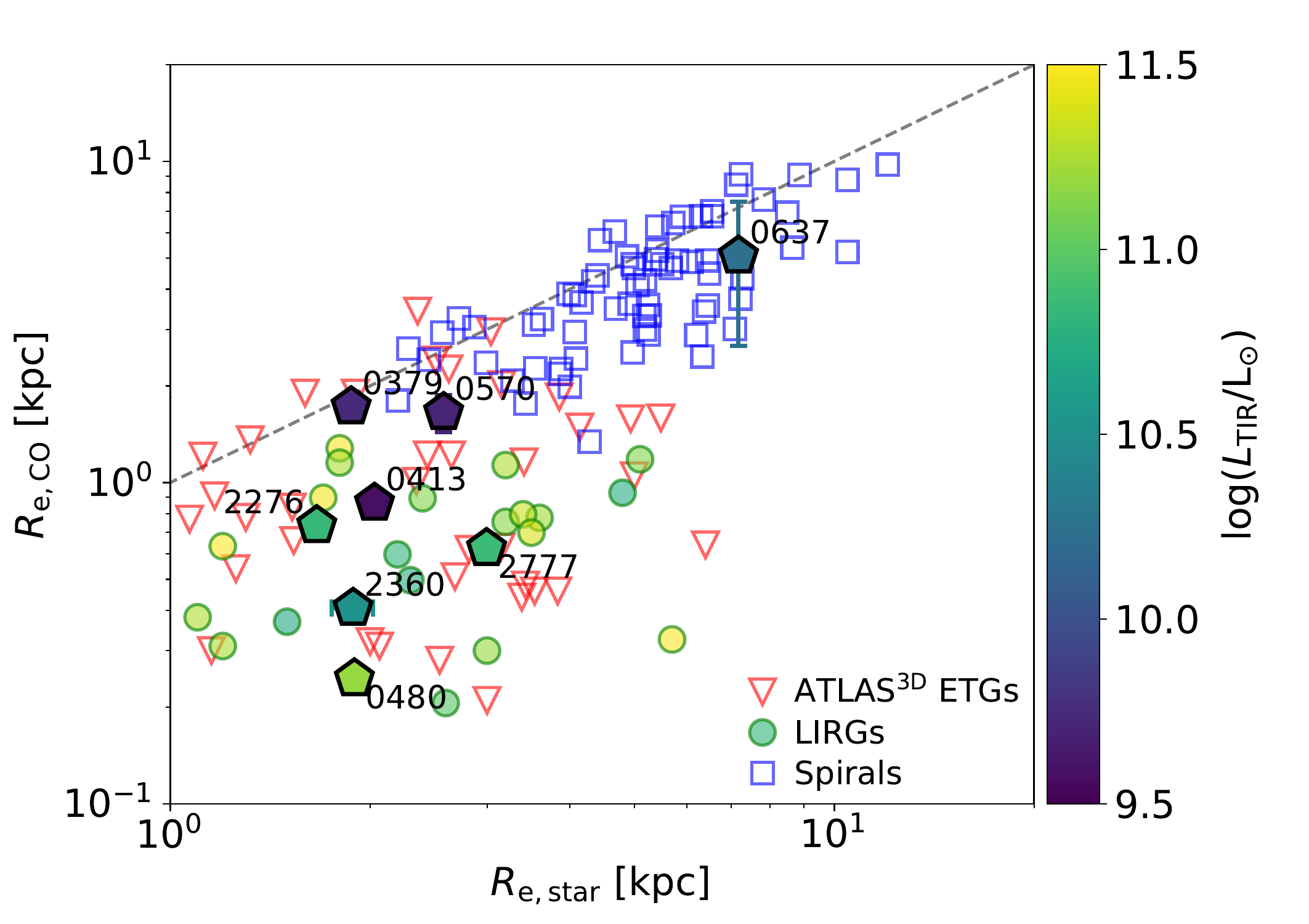}
\includegraphics[width=0.49\linewidth]{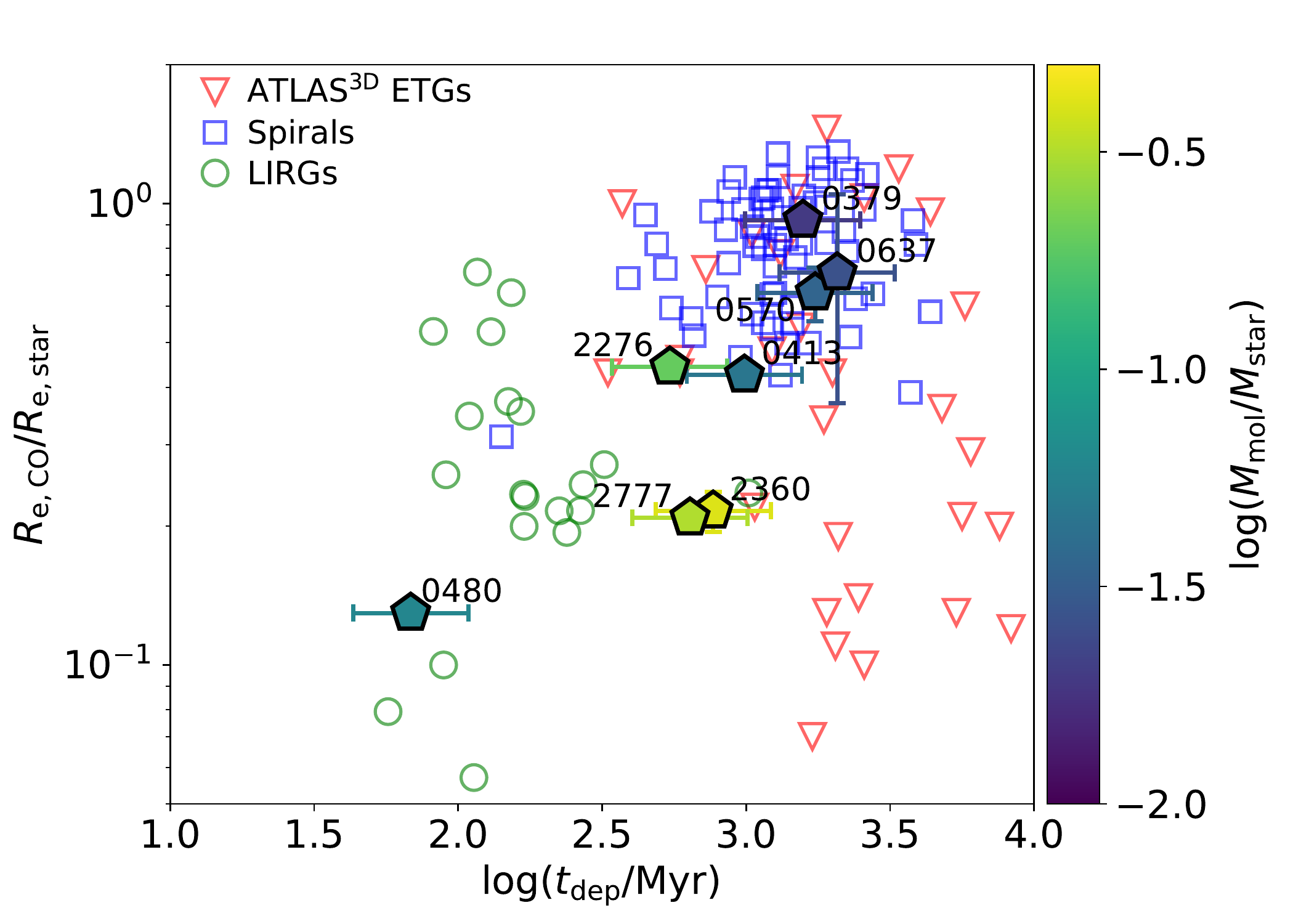}
\vspace{-8pt}
\caption{\textit{Left}: Effective radii of CO emission (gas) versus stellar component measured for PSBs in our sample, colour-coded with their IR luminosities (pentagons with black edges). 
Local early-type galaxies (open red triangles, CO sizes are {isodensity radii at $\sim10$\,\si{M_\odot.pc^{-2}} and thus likely larger than $R_\mathrm{e}$;} \citealt{davis13}), LIRGs (circles with green edges; \citealt{bellocchi22}) and spiral galaxies (open blue squares; \citealt{bolatto17}) are plotted for comparisons.
Dashed grey line denotes $R_\mathrm{e,CO} = R_\mathrm{e,star}$.
\textit{Right}: Gas-to-stellar size ratio versus gas depletion time scale. 
PSBs in our sample are colour-coded with their molecular gas fraction, {and the uncertainty of $t_\mathrm{dep}$ is assumed as 0.2\,dex for all sources. }
}
\vspace{-8pt}
\label{fig:03_size}
\end{figure*}

We first extracted 1.3\,mm continuum flux densities of eight PSBs in all the continuum maps generated with various angular resolutions.
Among them, we identified secure continuum emission associated with five sources (0413, 0480, 0637, 2276 and 2777) through a comparison between the signal-to-noise (S/N) ratios of positive and negative peaks in the maps.
In order to recover potentially diffuse emission and achieve an optimal S/N, continuum flux densities are typically measured in the 80\,\si{\kilo\lambda}-tapered maps (beam size $\sim$1\farcs5) through image-plane fitting (\textsc{casa/imfit}) except for 0413, which was only observed with a 5\arcsec\ beam. 
Among them, 0480 and 2777 exhibit strong continuum emission ($S_\mathrm{1.3mm}>4$\,\si{mJy}) as reported in \citetalias{smercina22}, {although our measured flux densities are significantly different ($4.11\pm0.05$ and $4.72\pm0.24$\,mJy in this work  versus $7.54\pm0.22$ and $1.00\pm0.35$\,mJy in \citetalias{smercina22}, respectively)}.
{Compared with \citetalias{smercina22}, the better recovery of extended emission with \textit{uv}-tapering resulted in the increase of measured $S_\mathrm{1.3mm}$ of 2777. 
For 0480, the decreased $S_\mathrm{1.3mm}$ suggests no excess of 1.3\,mm emission to the best-fit spectral energy distribution model, in contrast to the findings in \citetalias{smercina22}.
}
The continuum of 0637 was reported as a non-detection in \citetalias{smercina22} ($<$0.45\,mJy), but is detectable in the \textit{uv}-tapered map ($\mathrm{S/N}=4.7$, $0.86\pm0.25$\,mJy), indicating its extended nature.
0413 and 2276 were not included in \citetalias{smercina22}, but both are detectable in continuum emission.

To recover potentially extended CO emission, we extracted CO\,(2--1) spectrum for each source at the peak pixel of the line cube with the largest beam size {in which the source is unresolved.
These spectra are generally consistent with those extracted from 250k$\lambda$-tapered cubes with sufficiently large apertures ($r=6$\arcsec), but the aperture-extracted spectra are much higher in RMS noise.
}
We fitted the extracted CO\,(2--1) spectra with Gaussian profiles to determine the central frequencies and line widths (full-width at half maximum, FWHM), and constructed moment\,0 maps (integrated line intensity) within the velocity range of ($-$FWHM, $+$FWHM) from the line centre ($=$\,2$\times$ FWHM window) using \textsc{casa/immoments}. 
Moment\,1 and 2 maps (velocity field/dispersion) were then computed for regions at $\mathrm{S/N}\geq 3$ in the moment\,0 maps through a pixelated spectral fitting routine assuming Gaussian profile.
Figure~\ref{fig:01_spec} displays the ALMA CO\,(2--1) moment\,0 and 1 maps, spectra and continuum maps for all eight PSBs in our sample.
The CO\,(2--1) emission of all the PSBs is well resolved as visualised by the velocity gradients and comparison between the sizes of beams and $3\sigma$ (the lowest-level) contours in the moment\,0 maps.

We further modelled the surface brightness profiles of the CO\,(2-1) emission {over the 2$\times$FWHM spectral windows} in the \textit{uv}-plane. 
We first shifted the phase centres of all line measurement sets to the peak coordinates of the CO emission, using \textsc{casa/fixvis}, and extracted the circularised visibility profile (the real part of the amplitude) in \textit{uv}-distance bins from $10^{0.5}$ to $10^{3.5}$\,\si{\kilo\lambda} (the bin size is 0.1\,dex) using \textsc{casa/visstat}.
We then fitted the extracted visibility profiles with the Fourier transform of the 2D exponential, Gaussian and uniform disc functions:
\begin{flalign}
A_\mathrm{Exp}(x) = f x_0^2 / (x_0^2 + x^2) & \quad \Rightarrow \quad R_\mathrm{e} = 55.1/x_0 \\
A_\mathrm{Gaussian}(x) = f \exp(-x^2/x_0^2)
& \quad \Rightarrow \quad
R_\mathrm{e} = 54.7/x_0 \\
A_\mathrm{disc}(x) = f \sin(\pi x / x_0)/(\pi x / x_0)
& \quad \Rightarrow \quad
R_\mathrm{e} = 72.9/x_0
\label{eq:exp}
\end{flalign}
where $x$ is the \textit{uv}-distance, $x_0$ is the \textit{uv}-plane scale length, both in the unit of \si{\kilo\lambda}, and $f$ is the total line flux. 
Circularised effective (half-light) radius $R_\mathrm{e}$ can be derived from $x_0$.
Given the linearity properties of Fourier transform, we also fitted the \textit{uv}-profile with combinations of simple models (e.g., Gaussian$+$Exponential) up to three different components.
The visibility profiles and best-fit models are shown in Figure~\ref{fig:02_vis}.

Among the eight PSBs in our sample, six sources (0480, 0570, 0637, 2276, 2360, and 2777) exhibit a smaller reduced $\chi^2$ 
when their visibility profiles are fitted with models consisting of two components of different scale lengths. The extended CO components of these sources typically have effective radii of $R_\mathrm{e} = 1.6_{-0.7}^{+3.5}$\,kpc, tracing the gas component in the galaxy disc.
They are $1.4_{-0.3}^{+0.6}$ times brighter {(in terms of integrated luminosity)} than the inner compact CO components with $R_\mathrm{e} = 0.4_{-0.2}^{+0.4}$\,kpc, which trace the circumnuclear gas. 
0379 is best fitted with a single uniform disc model, and 0413 is best fitted with a single exponential disc model. 
We note, however, that 0413 has not been observed with the 12\,m array, and therefore the potential existence of a compact CO component cannot be ruled out.

We measured the effective radii through the integration of the image-plane models whose parameters were determined from the \textit{uv}-plane modelling, and a median $R_\mathrm{e,CO}$ of $0.8_{-0.4}^{+0.9}$\,kpc was derived.
The CO line fluxes were also modelled in the \textit{uv}-plane, which are $1.2_{-0.1}^{+0.7}$ times higher than that extracted at the peak pixel of the CO line cubes.
The CO\,(2--1) line luminosities were then calculated using the line fluxes and redshifts following \citet{solomon05}.
All these measurements are summarised in Table~\ref{tab:01_all}.

\vspace{-18pt}

\section{Discussion}
\label{sec:04_dis}

\subsection{Missing Flux Issue in \citet{smercina22}}
\label{ss:04a_comp}

Six PSBs in our sample have been initially analysed by \citetalias{smercina22}. 
Through image-plane fitting of the CO moment\,0 maps with a Gaussian profile, \citetalias{smercina22} reported that all six sources exhibit unresolved CO cores. 
This could be caused by the dominance of a compact CO component with a high surface brightness through image-plane fitting.
Without decreasing the weighting with long-\textit{uv}-distance (i.e., small-scale) data in the step of synthesised imaging (e.g., using larger \textsc{robust} parameters and \textit{uv}-tapering), the bulk of extended CO emission {can be} missed.
This is also visualised in Figure~\ref{fig:02_vis}, which shows poor fits of the \citetalias{smercina22} models to the observed visibility profiles.

On the other hand, the CO\,(2--1) line luminosities measured in this work utilising the visibility-plane analysis are $2.0_{-0.6}^{+3.1}$ times larger than the values reported by \citetalias{smercina22}.   
We note that the CO\,(2--1) line luminosities of the published single-dish measurements \citep{french15} are also $2.0_{-1.2}^{+1.7}$ times larger than those reported by \citetalias{smercina22} although with a larger uncertainty. 
This is consistent with the physical picture that extended CO components exist in gas-rich PSBs, which could be missed (i.e., resolved out) with interferometric observations with a high spatial resolution ($\lesssim 0.2$\,kpc).

\vspace{-12pt}

\subsection{Gas versus Stellar Sizes of PSBs}
\label{ss:04b_size}

The left panel of Figure~\ref{fig:03_size} shows the comparison of molecular gas sizes (traced by CO emission) versus stellar sizes of PSBs.
The effective radii of stellar components were measured in the SDSS \textit{r}-band \citep{sdssdr12} using \textsc{galfit} \citep{galfit} assuming S\'{e}rsic profiles, which are consistent with those reported in \citet{li19} measured as Petrosian half-light radii (median ratio is $0.95\pm0.04$).
The median gas-to-stellar size ratio is $0.43_{-0.21}^{+0.27}$ for PSBs in our sample.
{Although such a ratio does indicate a compact molecular gas reservoir, this value is} five times larger than that measured by \citetalias{smercina22} ($0.09_{-0.03}^{+0.06}$) and consistent with those of {fifteen} PSBs reported recently by \citet[][$0.38\pm0.24$]{baron22}, in which the CO sizes are measured with NOEMA through a single Gaussian profile fitting in the \textit{uv}-plane. 

We also compare the gas and stellar sizes of PSBs with those measured for LIRGs \citep{bellocchi13,bellocchi22}, spiral galaxies \citep{gd14,bolatto17} and gas-rich early-type galaxies (ETGs; \citealt{davis13}) in the local Universe.
At given stellar size, we find that the CO size distribution of PSBs are {comparable to} those of LIRGs and early-type galaxies.
The stellar and gas sizes of most extended source in our sample (0637) resemble those of spiral galaxies.

We also show that the gas-to-stellar size ratios of PSBs have {a} {potential} dependence on their gas depletion time scales ($p$-value\,$=$\,0.02 from Spearman's ranking coefficient) in the right panel of Figure~\ref{fig:03_size}.
We compute molecular gas mass ($M_\mathrm{mol}$) from the CO\,(1--0) luminosity \citep[measured by][]{french15} assuming a Galactic CO--to--H$_{2}$ conversion factor of $\alpha_\mathrm{CO} = 4.35$\,\si{M_\odot.(K.km.s^{-1}.pc^{2}.)^{-1}}, following \citetalias{smercina22}.
This is equivalent to the estimate using ALMA CO\,(2--1) luminosity assuming a $L^\prime_\mathrm{CO(2-1)}/L^\prime_\mathrm{CO(1-0)}$ ratio ($R_{21}$) of $0.9\pm0.1$, which is frequently seen in the LIRG environment \citep[e.g.,][]{papadopoulos12}.
The star-formation rate is derived from both the near-ultraviolet and total IR luminosities reported in \citet{smercina18} with the conversions in \citet{ke12} based on the \citet{kroupa01} initial mass function (see further discussion in Section~\ref{ss:04c_sf}).
The gas depletion time scales ($t_\mathrm{dep} = M_\mathrm{mol} / \mathrm{SFR}$; {assuming a typical uncertainty of 0.2\,dex from empirical calibrations}) are $10^{2.9\pm0.2}$\,\si{\mega yr} for sources in our sample, which are between those of LIRGs ($10^{2.1\pm0.2}$\,\si{\mega yr}) and spiral ($10^{3.1\pm0.3}$\,\si{\mega yr}) or early-type galaxies ($10^{3.2\pm0.6}$\,\si{\mega yr}).

The gas depletion time scale is known as a quantitative indicator of the age of star-forming galaxies.
Specifically, the post-starburst age of PSBs is found to be correlated with $t_\mathrm{dep}$ \citep{li19}. 
We note, however, that such a relation can be less conspicuous when the far-IR SFRs are used instead of optical (e.g., H$\alpha$) SFRs \citep{baron22}.
The {potential} dependence of gas-to-stellar size on $t_\mathrm{dep}$ places gas-rich PSBs as transitional objects between LIRGs and spiral galaxies (and further early-type galaxies with longer $t_\mathrm{dep}$).
During the ageing of PSBs, the compact circumnuclear gas component undergoes a faster depletion when compared with the extended disc component because of stronger stellar/supernova feedback \citep{sell14} or AGN feedback \citep{hopkins06}, leading to the broadening of gas profile and increased gas-to-stellar size ratio as observed.
Such an ``inside-out'' quenching mechanism has been shown or proposed for massive gas-rich systems in the local Universe \citep[e.g.,][]{gd15} and at high redshift \citep[e.g.,][]{zolotov15,sun21a}.
The gas-to-stellar size ratios of PSBs in our sample are also anti-correlated with molecular gas fraction ($M_\mathrm{mol}/M_\mathrm{star}$) as shown with the colour-coding in the right panel of Figure~\ref{fig:03_size} ($p$-value\,$=$\,0.02), further reinforcing the statement that CO-extended PSBs are likely at a later evolutionary stage.

{We note, however, that the robustness of $R_\mathrm{e,CO}/R_\mathrm{e,star} - t_\mathrm{dep}$ dependence of PSBs is subject to small number statistics. 
Considering the typical uncertainty of $t_\mathrm{dep}$ and measured uncertainty of source size ratio in this work through Monte Carlo simulations, the median $p$-value of Spearman's ranking coefficient is 0.036, with 56\% of the realisations below 0.05.
Further high-resolution ALMA/NOEMA CO observations of low-redshift PSBs are needed for the validation of this dependence.
}
{We also note that a similar trend cannot be found between gas-to-stellar size ratios and post-starburst ages derived in \citet{french18} for sources in this sample.
We argue that the post-starburst age estimated with SDSS fibre spectroscopy could be biased towards the centres of the galaxies (see spatial variations of stellar ages reported in \citealt{chen19}) because the stellar FWHMs of all sources in the sample are larger than the fibre diameter (3\arcsec), in contrast to the global properties as probed by $t_\mathrm{dep}$ and $R_\mathrm{e,CO}/R_\mathrm{e,star}$ here.
Also, no infrared/millimetre information was included in the modelling of \citet{french18} for these gas-rich PSBs.
Further spatially resolved studies of star-formation history and efficiency are keys to disentangling the structural evolution history of PSBs.
}

\begin{figure}
\centering
\includegraphics[width=.97\linewidth]{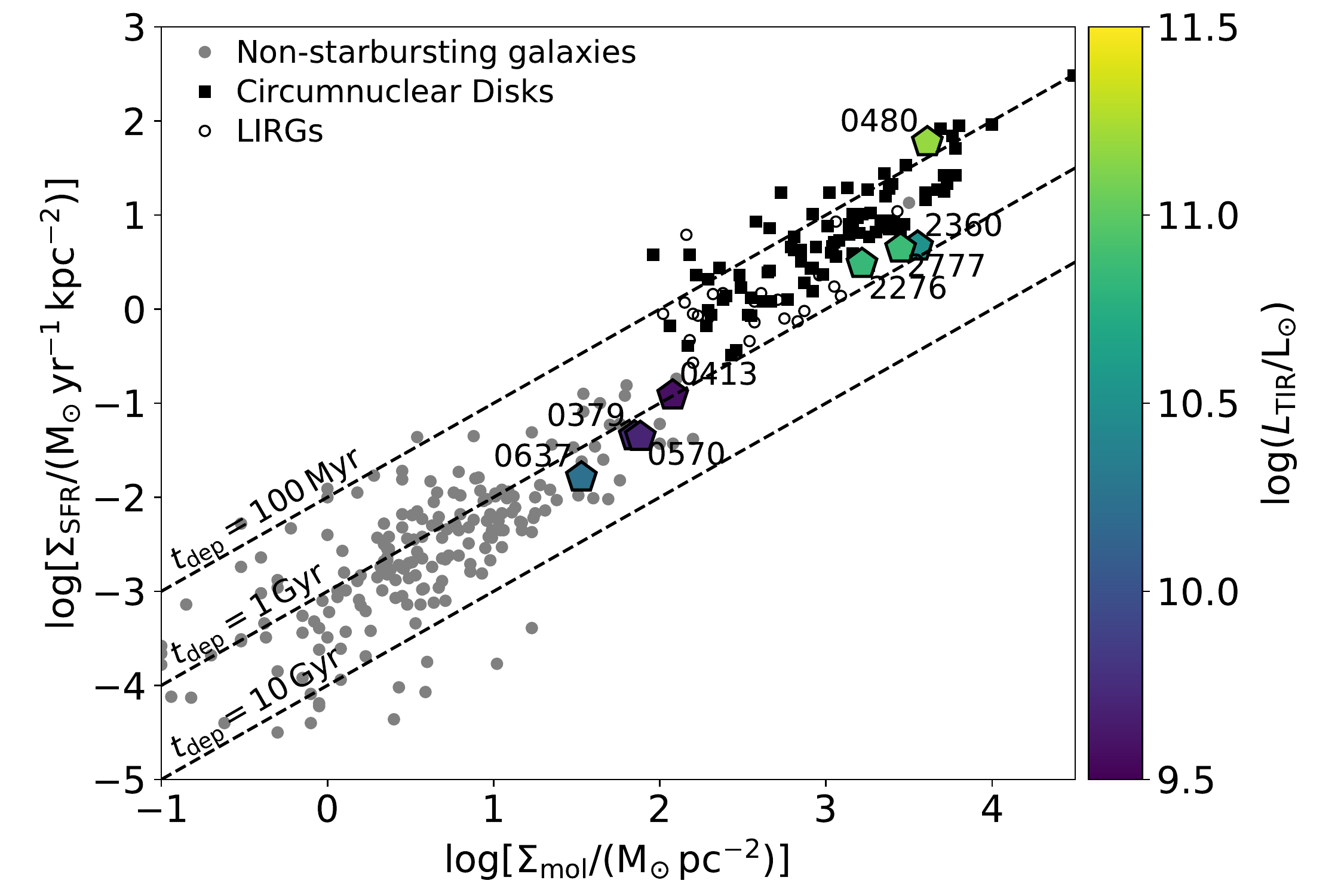}
\vspace{-8pt}
\caption{Surface density of SFR versus molecular gas mass of PSBs in our sample (pentagons), colour-coded with IR luminosity. 
Dashed lines denote constant depletion time scales of 0.1--10\,Gyr.
Non-starbursting galaxies \citep{dlr19}, circumnuclear discs and LIRGs \citep{kennicutt21} are plotted for comparisons.
}
\label{fig:04_ks}
\vspace{-8pt}
\end{figure}

\vspace{-12pt}

\subsection{Star-forming Activity in PSBs}
\label{ss:04c_sf}

\citetalias{smercina22} suggested that turbulent heating leads to a lower star-forming efficiency in PSBs when compared with galaxies with similar surface gas densities ($\Sigma_\mathrm{mol} = M_\mathrm{mol} / (2\pi R_\mathrm{e}^2)$) on the Kennicutt-Schmidt relation \citep{kennicutt98}.
We follow the same approximation of turbulent pressure ($P_\mathrm{turb} k_\mathrm{B}^{-1}$; \citealt{sun18}) for individual star-forming clouds used in \citetalias{smercina22}, and also consider a 0.3-dex overestimate of $P_\mathrm{turb}$ due to the beam smearing effect. 
With the larger CO sizes and therefore lower $\Sigma_\mathrm{mol}$ derived in this work, we estimate an average turbulent pressure of $\log[P_\mathrm{turb} k_\mathrm{B}^{-1}/(\mathrm{K}\,\mathrm{cm}^{-3})]=7.1\pm1.0$ for the PSBs in our sample
with the highest pressure of $10^{8.8}$\,\si{K.cm^{-3}} for 0480. 
These values are significantly lower than those reported by \citetalias{smercina22} and comparable to that measured for the Antennae, a major-merger starburst galaxy \citep{sun18}, meaning that 
the newly derived $P_\mathrm{turb}$ values are not high enough to motivate a suppression of SFR due to {turbulence in the} ISM.
Also, the beam sizes of the ALMA data analysed here 
(0.2--1.5\,kpc) are larger than the typical size of giant molecular clouds ($\lesssim$100\,pc), making the derived $P_\mathrm{turb}$ values somewhat uncertain.

Figure~\ref{fig:04_ks} shows the surface density of SFR ($\Sigma_\mathrm{SFR}$) versus molecular gas mass ($\Sigma_\mathrm{mol}$) on the Kennicutt-Schmidt relation \citep{kennicutt98}.
The CO sizes are used for the size approximation of the star-forming regions. 
As a whole, PSBs generally follows the $\Sigma_\mathrm{SFR} - \Sigma_\mathrm{mol}$ relation as those of star-forming galaxies  \citep[][]{kennicutt21}.
Therefore, it is not necessary to introduce turbulent heating as the physical process to suppress the star-forming efficiency in PSBs, consistent with the conclusion in \citet{baron22}.

This conclusion, however, depends on the reliability of IR luminosities as an SFR indicator for PSBs, which has been questioned by some recent studies.  For example, the use of $L_{\rm IR}$-based SFRs may lead to overestimates if there is a significant contribution from diffuse dust emission in ISM (e.g., \citealt{hayward14}).  In fact, Spitzer/IRS mid-IR spectroscopy of four PSBs in this sample shows that 
Neon-based SFRs will lead to a 0.4-dex decrease in $\Sigma_\mathrm{SFR}$ \citep{smercina18}.
On the other hand, \citet{baron22} showed that the use of $L_{\rm IR}$-based SFRs is probably valid with PSBs in the sample of \citet{french15}, the parent sample of the sources in this study. At this point, it is not yet clear if this uncertainty would be large enough to affect our conclusion.

\vspace{-12pt}

\section{Conclusions}
\label{sec:05_sum}

We present the analysis of the archival ALMA CO\,(2--1) and 1.3\,mm dust continuum data of eight gas-rich PSBs in the local Universe, six of which were studied by \citet{smercina22}. 
In contrast to this study reporting the detections of extraordinarily compact gas reservoirs in the six PSBs, we show that the molecular gas contents of these PSBs are clearly resolved in the \textit{uv}-plane, exhibiting effective radii of $R_\mathrm{e,CO} = 0.8_{-0.4}^{+0.9}$\,kpc.
The total CO\,(2--1) line fluxes are twice of those reported in \citetalias{smercina22}, making our values more in line with the published single-dish measurements.

We find that gas-rich PSBs typically consist of gaseous reservoirs at both the compact circumnuclear scale ($R_\mathrm{e}=0.4_{-0.2}^{+0.4}$\,kpc) and extended disc scale ($R_\mathrm{e}=1.6_{-0.7}^{+3.5}$\,kpc). 
The gas-to-stellar size ratios of PSBs in this sample are $0.43_{-0.21}^{+0.27}$, suggesting that the molecular gas components of PSB are {compact, while the ratios are comparable to} those of local LIRGs and early-type galaxies.
The gas-to-stellar size ratios of PSBs are {potentially} correlated with the gas depletion time scales, placing them as transitional objects between the LIRGs and early-type galaxies from an evolutionary perspective.
Finally, we find no sign of suppressed star formation in PSB from turbulent heating, and their surface densities of SFR and molecular gas are consistent with those of star-forming galaxies on the Kennicutt-Schmidt relation.

\vspace{-12pt}

\section*{Acknowledgements}
We thank the authors of \citetalias{smercina22} (Adam Smercina, J.~D.\ Smith, Decker French, and George Privon)
for helpful discussions.
{We thank the anonymous referee for helpful and constructive comments.}
FS acknowledges support from the NRAO Student Observing Support (SOS) award SOSPA7-022.
FS and EE acknowledge funding from JWST/NIRCam contract to the University of Arizona, NAS5-02105.
This paper makes use of the following ALMA data: ADS/JAO.ALMA\#2015.1.00665.S and 2016.1.00980.S. 
ALMA is a partnership of ESO (representing its member states), NSF (USA) and NINS (Japan), together with NRC (Canada), MOST and ASIAA (Taiwan), and KASI (Republic of Korea), in cooperation with the Republic of Chile. The Joint ALMA Observatory is operated by ESO, AUI/NRAO and NAOJ.
The National Radio Astronomy Observatory is a facility of the National Science Foundation operated under cooperative agreement by Associated Universities, Inc.

\vspace{-12pt}

\section*{Data Availability}

ALMA data used in this study can be download from the ALMA Science Archive hosted by each ALMA Regional Centre.
Other data can be shared upon reasonable request to the corresponding author.


\vspace{-12pt}


\bibliographystyle{mnras}
\bibliography{00_main} 






\bsp	
\label{lastpage}
\end{document}